\documentclass{revtex4-2}
\usepackage{amsmath,amssymb}  
\usepackage{amsfonts} 
\usepackage{graphicx,tikz,pgfplots} 
\usepackage[justification=centering]{caption}
\usepackage{array,booktabs}
\usepackage{adjustbox}
\usepackage{natbib}
\usepackage{url}
\usepackage[colorlinks,allcolors=blue]{hyperref}
\pgfplotsset{compat=1.14}
\usetikzlibrary{calc}
\usepgfplotslibrary{fillbetween}
\usepackage{tikz,comment,graphics,mathtools,siunitx,pgfplots}
\usetikzlibrary{calc,patterns,angles,quotes}
\usepackage[utf8]{inputenc}
\usepackage{amsthm}
\usetikzlibrary{angles, quotes}

\usepackage[english]{babel}
\usepackage[utf8]{inputenc}
\usepackage{graphicx}
\usepackage{subcaption}
\usepackage{float}
\usepackage[colorinlistoftodos]{todonotes}
\usepackage{pgfplots}
\pgfplotsset{compat=1.17}
\usetikzlibrary{calc}
\usetikzlibrary{arrows.meta}
\pgfplotsset{compat=newest}
\begin{document}
\begin{abstract}
In this paper we show how students can measure optical features of smartphone displays through three experiments. Observing diffraction patterns from smartphone displays allows students  to determine the Pixels Per Inch (PPI). Observing reflections within a smartphone display provides information about touch glass thickness and pixel layer properties. Finally, water drops are used as miniature lenses to see the magnified image of the pixels beneath.  An enhanced theoretical model that covers both small and large droplets is provided.
\end{abstract}

\title{From Pixels to Patterns: Decoding Smartphone Display Properties through Diffraction, Reflection, and Refraction}

\author{Mamatha Ramanjineyulu Maddur$^1$}
  \author{Hemansh Shah$^2$}%
 
 \author{Praveen Pathak$^1$}%
 \email{praveen@hbcse.tifr.res.in}

 \affiliation{$^1$Homi Bhabha Centre for Science Education-TIFR\\Mankhurd, Mumbai, India}

 \affiliation{$^2$Indian Institute of Science, Bangalore, India}

\maketitle
\section{Introduction}
\begin{figure}[h!]
    \centering
    \input{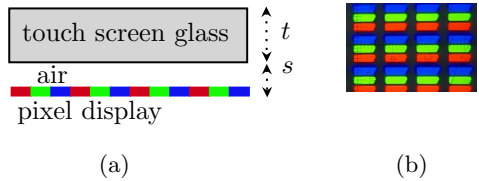}
    \captionsetup{justification=raggedright,singlelinecheck=false}
    \caption{(a) Schematic diagram of the display assembly of a phone consisting of a touch screen glass and RGB pixels. (b) magnified view of the pixel display. }
    \label{fig:screen_scheme}
\end{figure}

 Students spend a lot of time looking at their phones. Why not give them a new way to look at them? Here we explore three optical phenomena of typical smartphones, namely, diffraction, reflection, and refraction. A schematic diagram of the display unit is shown in Fig. \ref{fig:screen_scheme} where the pixels of the display unit are at a distance \(s\) away from the bottom of the glass surface, which has a thickness \(t\). 

Smartphone, laptop, and TV displays produce distinctive diffraction and interference patterns, as noted in earlier studies \cite{Jesus, kristin,Diffraction_jennifer}. In this work, we explore three related experiments using such displays: measuring pixel density (PPI) via diffraction, estimating touch glass thickness through reflection patterns, and refining water-drop lens analysis to account for thick lens effects. Given the short distances measured (typically close to a millimetre) in all the experiments, we used ImageJ \cite{ImageJ} for precise measurements.  The experiments are explained in detail in the following sections. 

\section{Diffraction Pattern of Pixels}

\begin{figure}[htbp]
    {
    \begin{minipage}[b]{0.22\textwidth}
        \includegraphics[width=\textwidth]{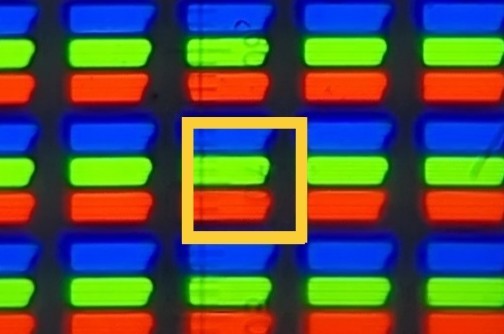}
        \subcaption*{(a)}
        \vspace{1em}
        \includegraphics[width=\textwidth]{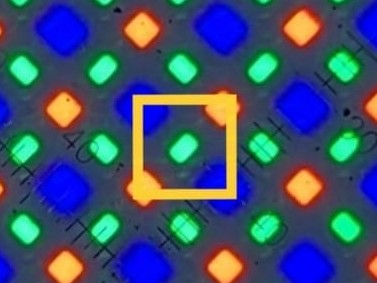}
        \subcaption*{(b)}
    \end{minipage}
    \hspace{1em}
    \begin{minipage}[b]{0.35\textwidth}
        \includegraphics[width=\textwidth]{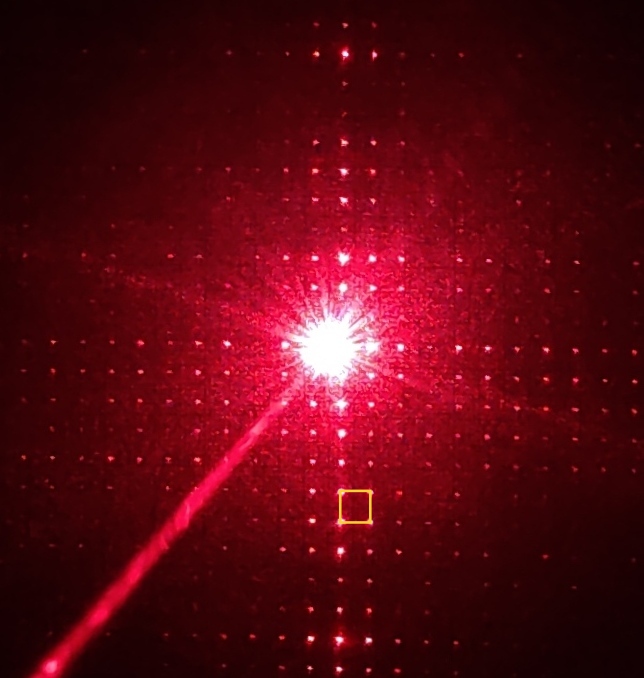}
        \subcaption*{(c)}
    \end{minipage}
    \hspace{1em}
    \begin{minipage}[b]{0.35\textwidth}
        \includegraphics[height=0.28\textheight,width =0.97\textwidth]{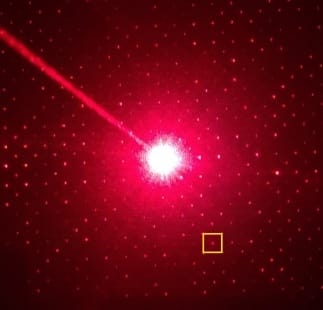}
        \subcaption*{(d)}
    \end{minipage}}
    \captionsetup{justification=raggedright,singlelinecheck=false}
    \caption{(a) Magnified image of the pixel arrangement in a Motorola G6 Play phone's display. 
(b) Magnified image of the pixel arrangement in a Google Pixel 6a phone's display. 
(c) Diffraction pattern of the Motorola G6 Play phone's display. 
(d) Diffraction pattern of the Google Pixel 6a phone's display. The yellow box in the magnified images depicts one pixel in the pixel array.}
    \label{fig:Motorola&Googlepixel}
\end{figure}

\begin{figure}
   { \centering
    \includegraphics[width=0.5\linewidth]{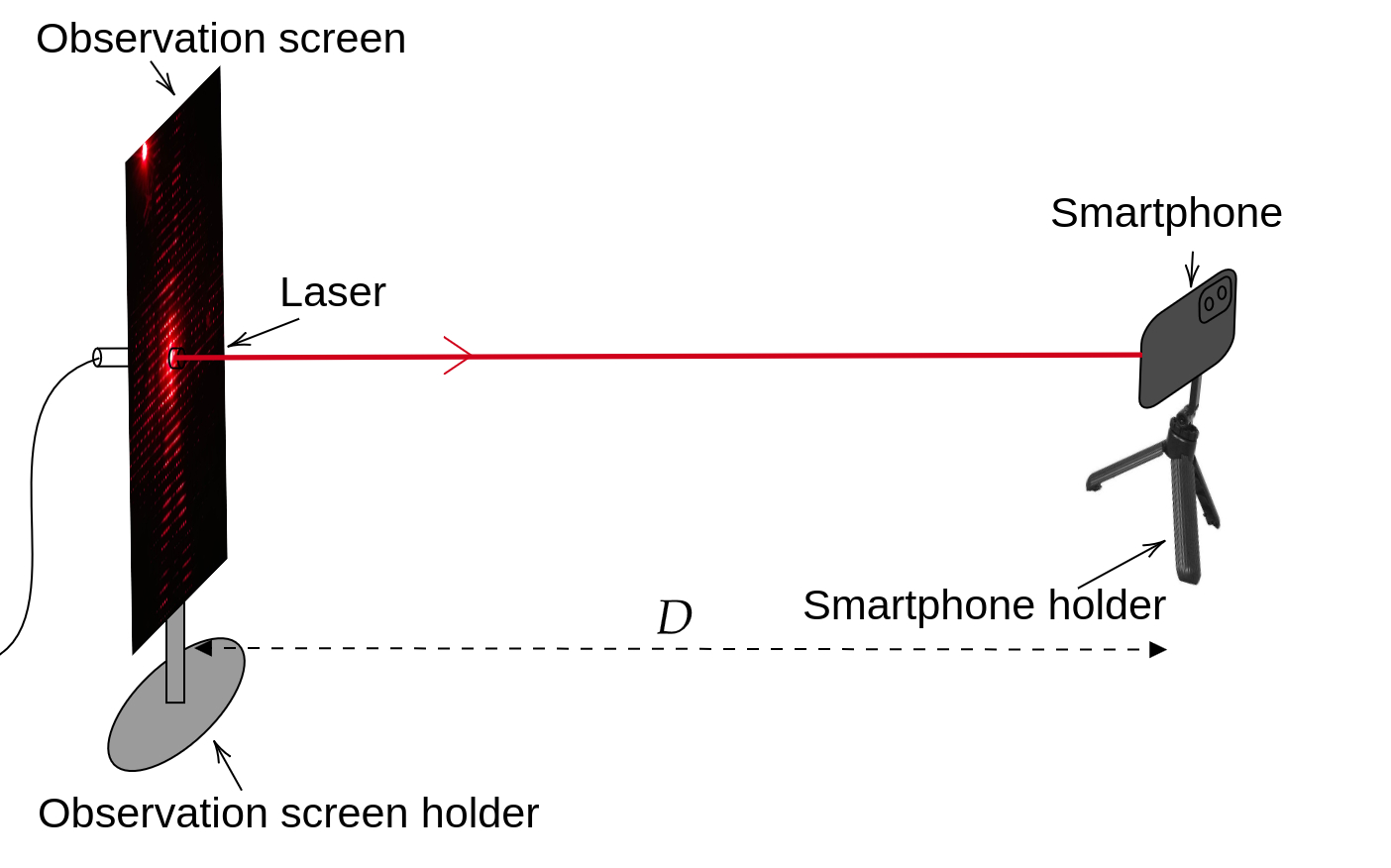}
    \caption{{Schematic diagram of the diffraction experiment set up (Sec. II).}}
    \label{fig:Diffraction_schematic_figure}}
\end{figure}

The pixels in a smartphone display are arranged periodically, with each pixel composed of red, green, and blue subpixels separated by black matrix padding~\cite{ian_pixel}. Magnified view, as seen through a microscope, of the phone's display is shown in Fig.~\ref{fig:Motorola&Googlepixel}(a) and (b). Here, the different colored regions differ in reflectivity and act as scattering centres.  When monochromatic light is reflected from such a structured surface, the abrupt changes in reflectivity at the subpixel edges lead to diffraction, resulting in interference patterns such as those shown in Fig.~\ref{fig:Motorola&Googlepixel}(c) and (d) respectively. The setup for this part is a simple setup where the laser light is incident near normally on the phone's display and the reflected light is observed on the observation screen (see Fig. \ref{fig:Diffraction_schematic_figure}).

\begin{figure}[h!]
    \includegraphics[scale=0.3]{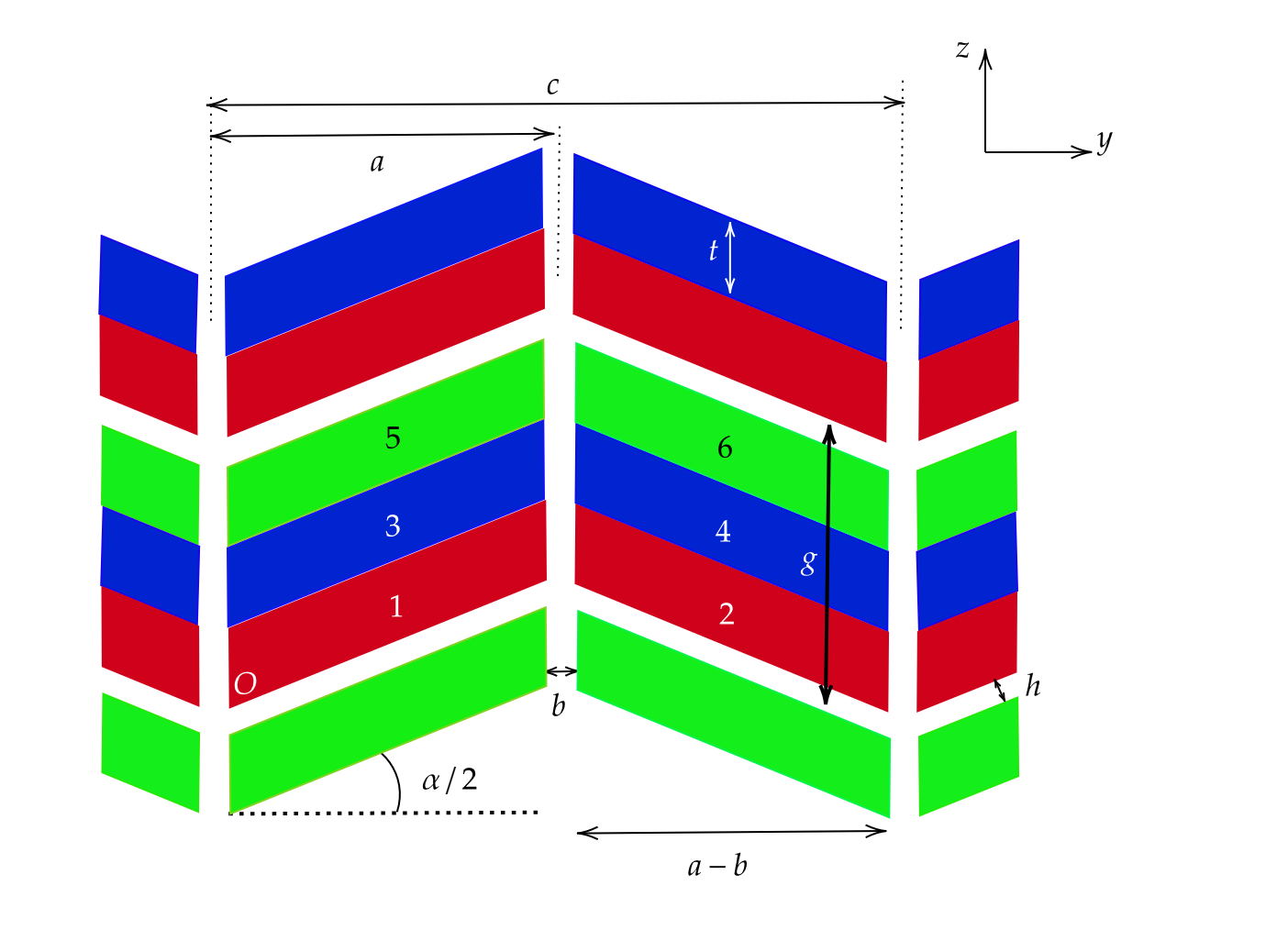}
\captionsetup{justification=raggedright,singlelinecheck=false}
    \caption{Schematic representation of the pixel array. Two arrays of RGB pixels are tilted at an angle $\alpha$ with respect to each other. The individual subpixels are labeled numerically (1 to 6). The spacing between adjacent subpixels is small and negligible compared to the subpixel thickness $t$ and the centre-to-centre distance between repeating pixel units or the pixel pitch $g$.  The inter-pixel separation between two neighbouring pixel units is $h$. The arrays are separated by a distance $b$ along the $y$-direction. All the subpixels have the same thickness $t$.}
    \label{fig:RGB}
\end{figure}

To understand this diffraction pattern in more detail, we consider a schematic diagram of the display. In Fig. \ref{fig:RGB}, two arrays of RGB pixels are placed at an angle with respect to each other. Certain smartphones feature such displays, one of which will be examined later in this section.  The individual subpixels are labelled numerically (1 to 6), and the origin is chosen at the corner of a red subpixel number 1. The relevant separations and geometric parameters are also indicated in the figure.

\begin{figure}[h!]
    \centering
    \includegraphics[width=0.5\linewidth]{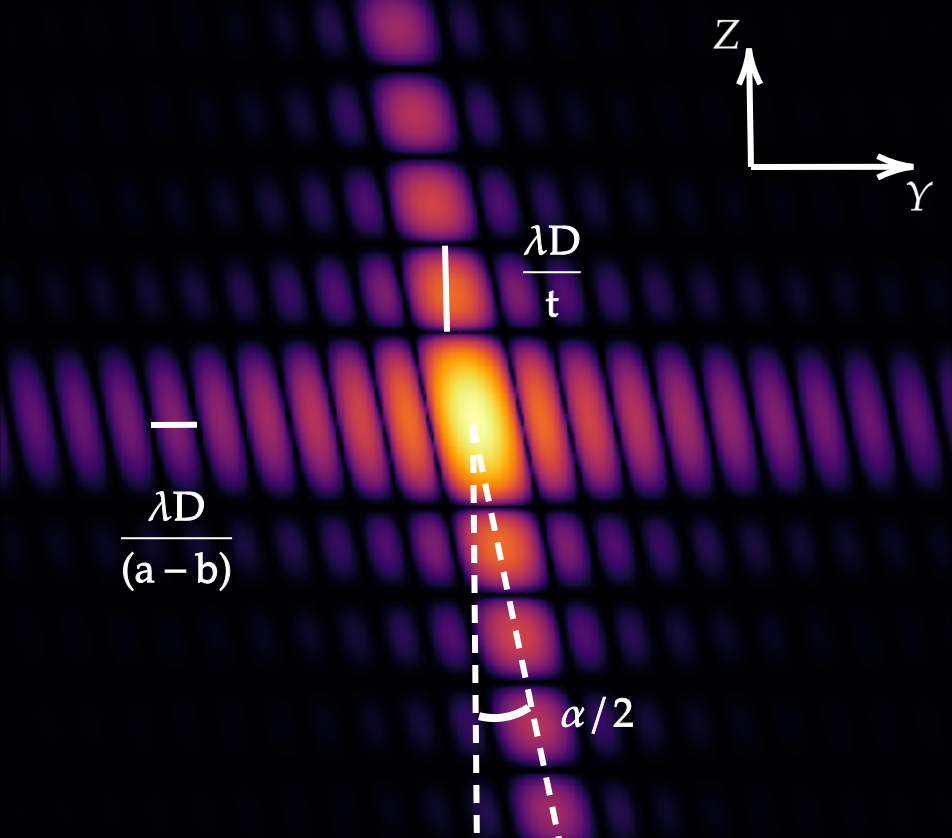}
    \caption{Simulated diffraction pattern due to a single red subpixel 1.}
    \label{fig:single_red_pixel_diff}
\end{figure}

The amplitude of the Fraunhofer diffraction field due to a single red subpixel 1 (see Fig.~\ref{fig:RGB}) at a screen placed a large distance $D$ away is  
\begin{align}
E_{1}(Y,Z) &\propto \int_{0}^{a-b} dy \int_{y\tan{(\alpha/2)}}^{y\tan{(\alpha/2)}+t}  
e^{ik\tfrac{Y}{D}y} 
e^{ik\tfrac{Z}{D}z}\,dz,
\end{align}
where $(y, z)$ and $(Y, Z)$ represent the aperture and screen coordinates, respectively, and $k = {2\pi}/{\lambda}$ is the wavenumber of the incident light. Evaluating the integral gives  
\begin{align}
E_{1}(Y,Z) &\propto (a-b) t \,
\operatorname{sinc}\!\left( \frac{(a-b)(Y + Z \tan (\alpha/2))}{\lambda D}\right) \,
\operatorname{sinc}\!\left( \frac{Z t}{\lambda D} \right)\label{eq:AR1},
\end{align}
where
\[
\operatorname{sinc}(x)=\frac{\sin(\pi x)}{\pi x}.
\]
The intensity is
\[
I_{1}(Y,Z)\propto E^2_{1}.\]
Function  $\operatorname{sinc}(x)$ has zeros at $x=\pm 1,\pm 2,\dots \pm n$. Along the $Y$-axis, the intensity will be minimum at
\[
{Y = -Z\tan(\alpha/2) + n\,\frac{\lambda D}{(a-b)}\; }.
\]
Similarly, along the $Z$-axis, the intensity will be minimum at
\[
{\;Z = n\,\frac{\lambda D}{t}}.
\]
A simulated diffraction pattern from a single rectangular subpixel (using Eq. (\ref{eq:AR1})) is shown in Fig. \ref{fig:single_red_pixel_diff}, with the locations of the minima indicated. The tilt observed in one arm of the diffraction pattern arises from the tilt of the subpixel 1.

Next, consider two adjacent red subpixels 1 and 2 in Fig.~\ref{fig:RGB}. The corresponding field is  
\begin{align}
E_{12}(Y,Z) \propto (a-b)t\,\operatorname{sinc}\!\left(\tfrac{Zt}{\lambda D}\right) 
&\Bigg[
e^{-\tfrac{i\pi Ya}{\lambda D}} \,
\operatorname{sinc}\!\left( \frac{(a-b)(Y + Z \tan (\alpha/2))}{\lambda D}\right) \nonumber \\
&\quad+ e^{\tfrac{i\pi Ya}{\lambda D}} \,
\operatorname{sinc}\!\left( \frac{(a-b)(Y - Z \tan (\alpha/2))}{\lambda D}\right)
\Bigg],
\end{align}
The diffracting field $E$ due to the repeating pixel unit is obtained by superposing the fields from all subpixels (1 to 6) shown in Fig.~\ref{fig:RGB}. Thus
\begin{align}
E &\propto \big( A_{R} e^{\tfrac{-i2\pi Z t}{\lambda D}} \;+\; A_B \;+\; A_{G} e^{\tfrac{i2\pi Z t}{\lambda D}} \big)\, E_{12},
\end{align}
where $A_R$, $A_B$, and $A_G$ are the reflectivities of the red, blue, and green subpixels. This field represents the combined contribution from one complete pixel element and serves as the basis for calculating the intensity distribution on the observation screen. Smartphone displays contain many such pixels arranged in a periodic grid. To account for this, we apply the {array theorem}~\cite{hecht_nslit}, which states that the total electric field due to an array of identical apertures is the product of the single-aperture diffraction envelope and the interference pattern from the array. Mathematically, for an array of identical apertures spaced at positions \( x_l \), the total field becomes:

\begin{align}
E_\text{total}(\theta) = E_\text{single}(\theta) \cdot \sum_l e^{i k x_l \sin\theta}.\label{eq:array}
\end{align}
Thus the overall diffracted intensity due to an array of $N$ repeating units along the $y$-axis and $M$ units along the $z$-axis   is given by
\begin{align}
I(Y,Z) &= \left[
(N+1)(M+1) \;
\frac{\operatorname{sinc}\!\left(\tfrac{Yc(N+1)}{\lambda D}\right)}{\operatorname{sinc}\!\left(\tfrac{Yc}{\lambda D}\right)} \;
\frac{\operatorname{sinc}\!\left(\tfrac{Zg(M+1)}{\lambda D}\right)}{\operatorname{sinc}\!\left(\tfrac{Zg}{\lambda D}\right)} \;
E
\right]^2. \label{eq:totalI}
\end{align}
Compared to a single pixel, here $N\times M$ dimensional array sharpens the diffraction pattern. However, the intensity of the principal maxima is modulated by the internal structure of the repeating unit.

The intensity along $Z=0$, can be written as 
\begin{align}
    I(Y,0)\propto \left[\operatorname{sinc}\left(\frac{(a-b)Y}{\lambda D}\right)\cos{\left(\frac{\pi Ya}{\lambda D}\right)}\frac{\operatorname{sinc}\left(\frac{Yc(N+1)}{\lambda D}\right)}{\operatorname{sinc}\left(\frac{Yc}{\lambda D}\right)}\right]^2.\label{eq:Z=0}
\end{align}
Similarly, the intensity along $Y=0$, for equal reflectivities of red, blue and green sub pixels can be written as 
\begin{align}
I (0,Z)\propto \left[
\;
 \;
\frac{\operatorname{sinc}\!\left(\tfrac{Zg(M+1)}{\lambda D}\right)}{\operatorname{sinc}\!\left(\tfrac{Zg}{\lambda D}\right)} \;
\operatorname{sinc}\!\left(\tfrac{Z(g-h\sec{(\alpha/2)})}{\lambda D}\right)
\right]^2.  \label{eq:Y=0}
\end{align}

\begin{figure}
    \centering
    \includegraphics[width=0.7\linewidth,angle=0]{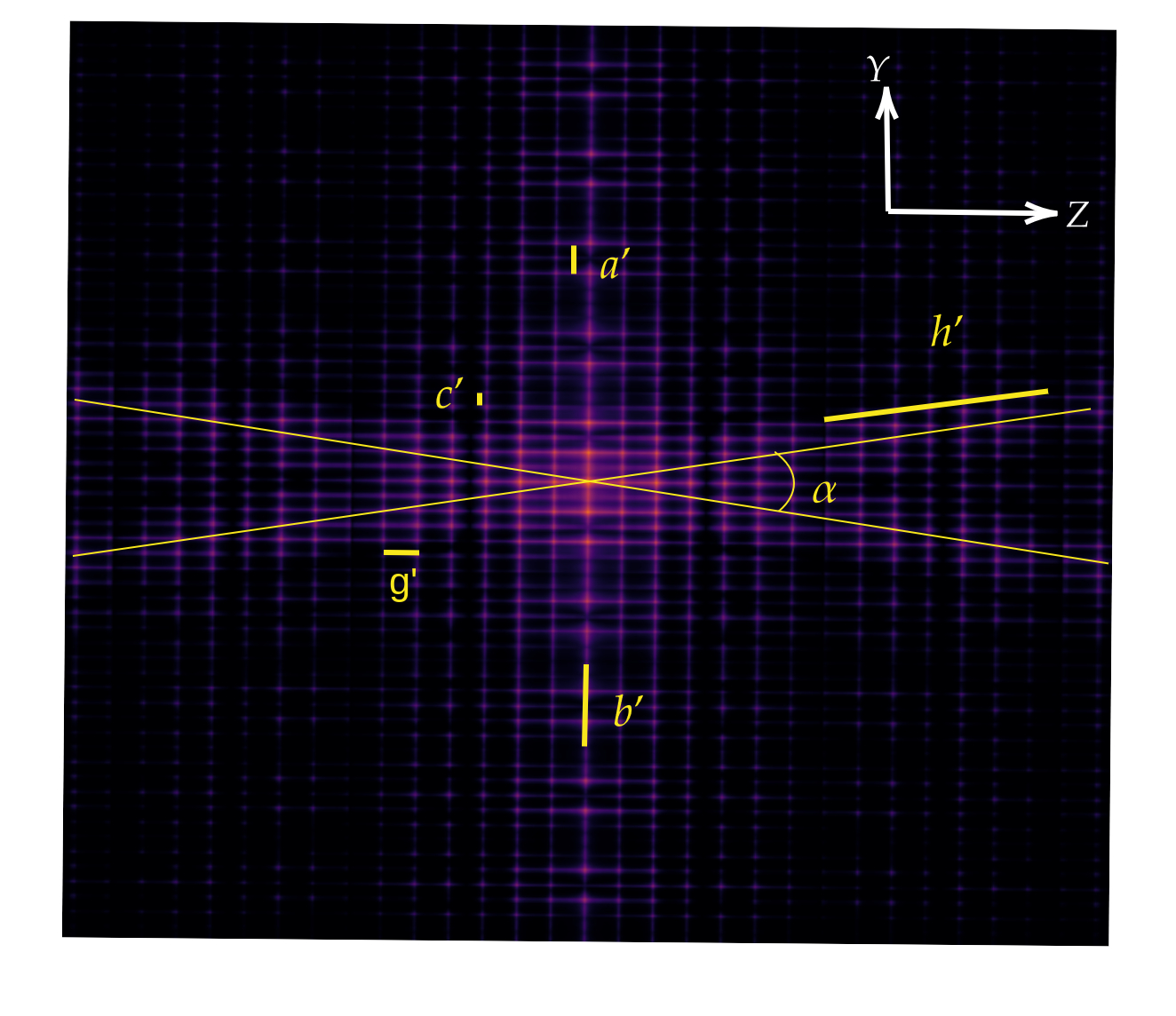}
    \caption{Simulated diffraction pattern due to pixels 1 to 6 of Fig. \ref{fig:Diffraction_schematic_figure}.}
    \label{fig:Full_Diff_small}
\end{figure}

Let us analyze Eqs.~(\ref{eq:Z=0}) and (\ref{eq:Y=0}). In Eq.~(\ref{eq:Z=0}),  along \(Z = 0\), the term  
\(
\cos^2\!\left({\pi Y a}/{\lambda D}\right)
\)
produces interference minima at \(Y = (2n+1){\lambda D}/{2a}\), where \(n\) is an integer. Since \(c = 2a\), every alternate maximum of the pattern produced by \(c\) coincides with a minimum of this cosine term. As a result, alternate bright spots are suppressed along the $Y$-axis of the diffraction image. The modulation arising from the subpixel width \(b\) comes from the factor \(\operatorname{sinc}^2[(a-b)Y/\lambda D]\), which defines the envelope of a single-subpixel diffraction pattern. This envelope modulates the sharper interference fringes due to the periodic spacing \(c\), giving rise to the intensity distribution observed along the $Y$-axis  of the pattern.  
The overall periodicity of the pixel array, represented by   
\[
\dfrac{\operatorname{sinc}^2\left(\tfrac{Yc(N+1)}{\lambda D}\right)}{\operatorname{sinc}^2\left(\tfrac{Yc}{\lambda D}\right)},
\]
determines the positions of the principal maxima, separated by \(\lambda D / c\). These diffraction intensity minima on the diffraction image on the  screen can be identified by  
 \(a'\), \(b'\), and \(c'\)  correspond respectively to the $a$, $b$, and $c$, and are related by \(a' = \lambda D/a\), \(b' = \lambda D/b\), and \(c' = \lambda D/c\). A theoretically obtained pattern from Eq. (\ref{eq:totalI}) is also shown in Fig.  \ref{fig:Full_Diff_small}, where the primed variables are depicted. 

Along the vertical direction (\(Y = 0\)), Eq.~(\ref{eq:Y=0}) shows that the periodic pixel rows, separated by a distance \(g\), generate bright fringes spaced by \(\lambda D / g\). The additional term \(\operatorname{sinc}^2[(g - h\sec(\alpha/2))Z / \lambda D]\) represents modulation due to the vertical offset \(h\) between subpixels tilted by an angle \(\alpha/2\). This modulation periodically weakens the maxima due to \(g\)  at intervals of \(h'\). This produces missing maxima along the line \(Y = \pm Z\tan(\alpha/2)\), which form the two arms of the X-shaped diffraction pattern. The corresponding spacings \(g'\) and \(h'\) observed on the diffraction image satisfy \(g' = \lambda D/g\) and \(h' = \lambda D/h\) (see Fig. \ref{fig:Full_Diff_small}). These relationships allow a direct correspondence between the structural parameters of the pixel array and the measurable features of its diffraction pattern.

We demonstrate the theory with three examples from different display technologies (IPS LCD, AMOLED and DotDisplay). Two representative cases (Moto G6 Play and Google Pixel 6a) are shown in Fig.~\ref{fig:Motorola&Googlepixel}. Panels (a) and (b) show magnified, microscope images of the pixel arrays, while (c) and (d) show the corresponding diffraction patterns obtained at near-normal incidence. The repeating unit is depicted by the yellow box. As the magnified view suggests that the pixels are not tilted, such that $\alpha=0$ in Eq. (\ref{eq:totalI}).

From the spacing between adjacent diffraction maxima (measured on the screen) and the known laser wavelength and geometry, we infer the pixel pitch using the grating relation \(\lambda D/x\).  The measured pixel densities (PPI) for the Motorola G6 Play and Google Pixel 6a are \( 287 \pm 2 \) and \( 431 \pm 4 \), respectively, in close agreement with the manufacturer's specified values of 282 and 429.

\begin{figure}[h!]
    \centering
    
    \begin{subfigure}[b]{0.95\textwidth}
        \centering
        \includegraphics[width=0.9\textwidth]{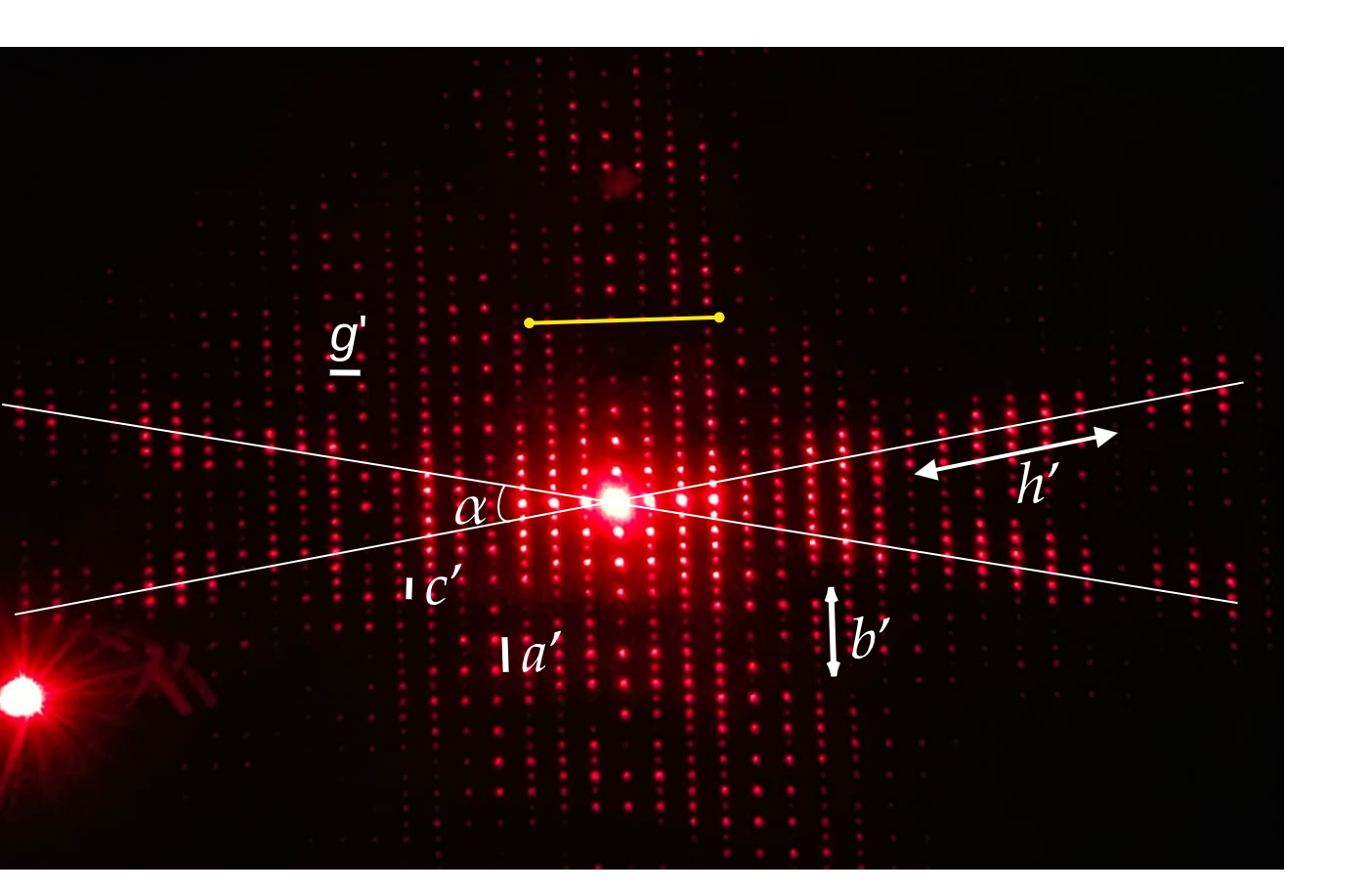}
        \caption{}
        \label{fig:a}
    \end{subfigure}
    \begin{subfigure}[b]{0.33\textwidth}
        \centering
        \includegraphics[width=0.9\textwidth]{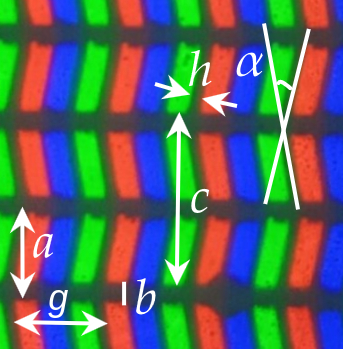} 
        \caption{}
        \label{fig:b}
    \end{subfigure}
    \begin{subfigure}[b]{0.55\textwidth}
        \centering
        \includegraphics[width=0.9\textwidth]{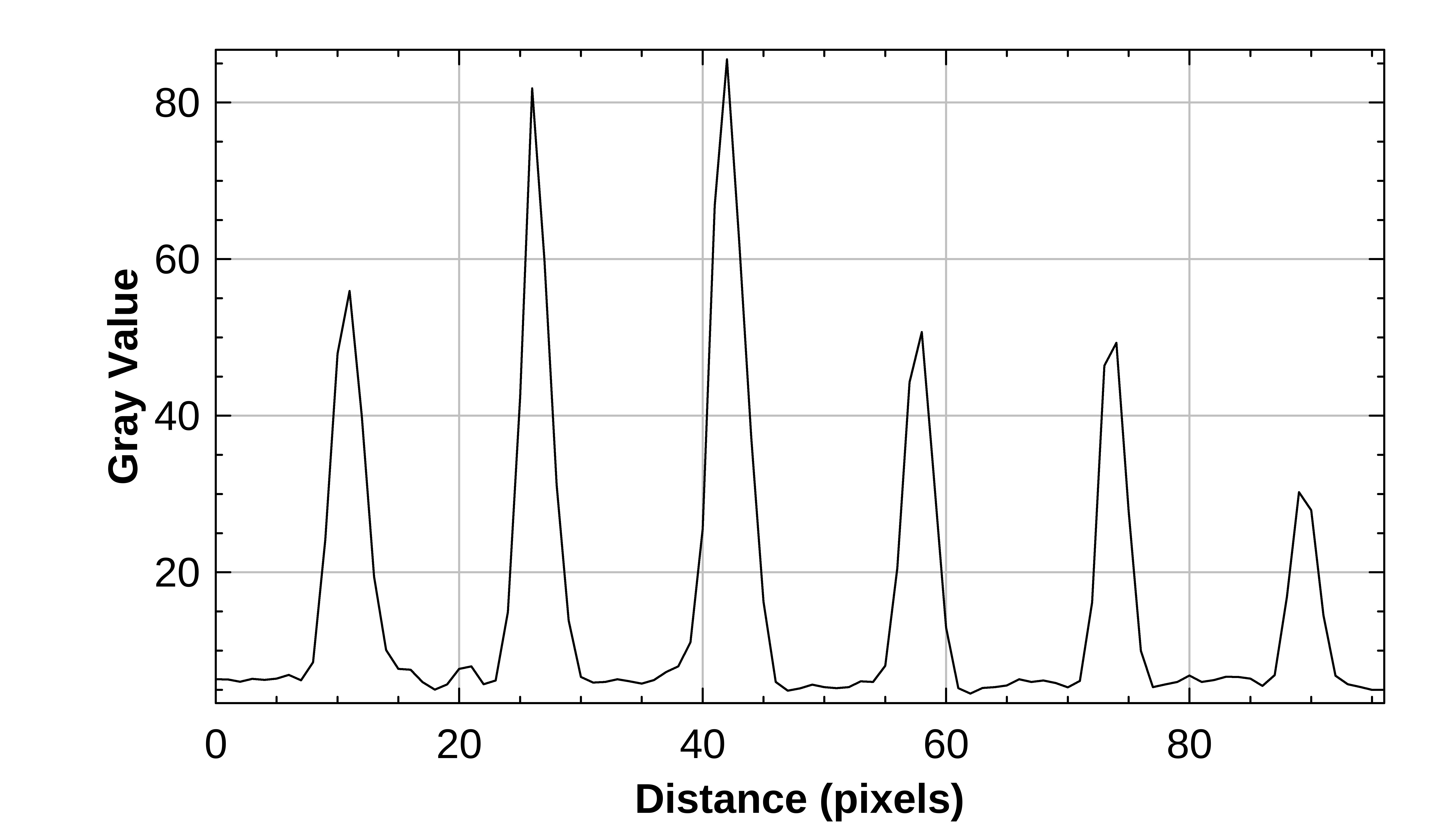} 
        \caption{}
        \label{fig:c}
    \end{subfigure}
    \captionsetup{justification=raggedright,singlelinecheck=false}
    \caption{(a) Diffraction pattern of the Redmi Note 9 Pro phone's display. (b) Phone's pixel pattern under a microscope. (c) The Gray value profile along the yellow line drawn on the pattern. This profile was obtained using ImageJ. {It is worth noting that not many phones have pixels aligned at a specific angle (denoted as $\alpha$ here). We observed this particular feature in a couple of phones with IPS LCD displays available to us.  The labels are defined in the text.}}
    \label{fig:diffraction}
\end{figure}

The Redmi Note 9 Pro has a pixel array in which neighboring pixel rows are tilted by an angle \(\alpha/2\). Its diffraction pattern is shown in Fig.~\ref{fig:diffraction}; the simulated pattern from Eq.~(\ref{eq:totalI}) is shown in Fig.~\ref{fig:Full_Diff_small} for comparison. Two notable features appear: (i) a set of sharp principal maxima corresponding to the pixel pitch, and (ii) an ``X'' shaped pattern whose arms are aligned at \(\pm\alpha/2\). As explained above, the principal maxima are spaced by \(\lambda D/c\), while the modulation in the X-shaped arms arises from the vertical offset \(h\) and tilt \(\alpha/2\) of the subpixel arrangement.

For this phone, the observation-screen distance was \(D=0.855\)~m. To resolve the small-scale structure associated with \(h\), we moved the screen closer when necessary. The primed distances \(a',b',c',g',h'\) visible on the diffraction image are mapped to the structural parameters \(a,b,c,g,h\) via \(x'=\lambda D/x\), and the measured values are summarized in Table~\ref{tab: Diffraction results}.

\begin{table}[h!]
\caption{Measurements of pixels patterns from diffraction and actual measurements of Redmi Note 9 Pro. }\label{tab: Diffraction results}
\begin{tabular}{lcc}
\hline\hline
Parameters & From the diffraction image & Using the microscope picture \\ \hline
$a$ (thickness of pixel array)      & $(62\pm1)\,\mu$m         & $(66\pm1)\,\mu$m      \\ 
$b$ (thickness of the padding between two pixel arrays)       & $(16\pm1)\,\mu$m         & $(18\pm1)\,\mu$m      \\ 
$c$  (height of the repeating pixel unit)    & $(123\pm3)\,\mu$m        & $(133\pm2)\,\mu$m     \\
$g$  (width of the repeating pixel unit)     & $(62\pm2)\,\mu$m         & $(68\pm2)\,\mu$m      \\ 
$h$  (thickness of the padding between the subpixels)    & $(12\pm1)\,\mu $m       & $(11\pm1)\,\mu$m     \\ 
$\alpha$ (angle between the subpixels of two adjacent arrays)  & $20^\circ \pm1^\circ$         & $19^\circ\pm2^\circ$      \\ \hline\hline
\end{tabular}

\end{table}

We measured distances on the diffraction images using ImageJ \cite{imageJ2}. A scale was provided by attaching graph paper to the observation screen. For each pattern, a line was drawn along the relevant row or arm of maxima, and the gray-value profile was extracted. As an example, the gray value profile along the yellow line in Fig.~\ref{fig:diffraction}(a) is shown in Fig.~\ref{fig:diffraction}(c). With this profile, the peaks (or minima, when peaks overlap) were located and the inter-peak separations averaged. Uncertainties reported in Table~\ref{tab: Diffraction results} represent one standard deviation from ten measurements. In our experience, measurements from diffraction images are often more precise than those from microscope photographs because the diffraction maxima are very sharp.

The experiments in the next two sections are done with the Redmi Note 9 Pro only.

\section{Reflections from the touch glass and the pixels}
\begin{figure}[htbp]
    \centering
    \begin{subfigure}[b]{0.51\textwidth}
        \centering
        \begin{tikzpicture}
            \node[inner sep=0pt] (img) at (0,0) {\includegraphics[height=6cm]{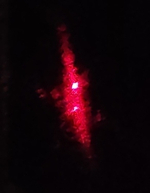}};
            \draw[yellow, thick] (-0.0,-0.5) -- (0.0,0.5); 
            \fill[yellow] (-0.0,-0.5) circle (1pt); 
            \fill[yellow] (0.0,0.5) circle (1pt);   
        \end{tikzpicture}
        \caption{}
        \label{fig:a}
    \end{subfigure}
    \hspace{-0.1\textwidth}
    \begin{subfigure}[b]{0.475\textwidth}
        \centering
        \includegraphics[height=6cm]{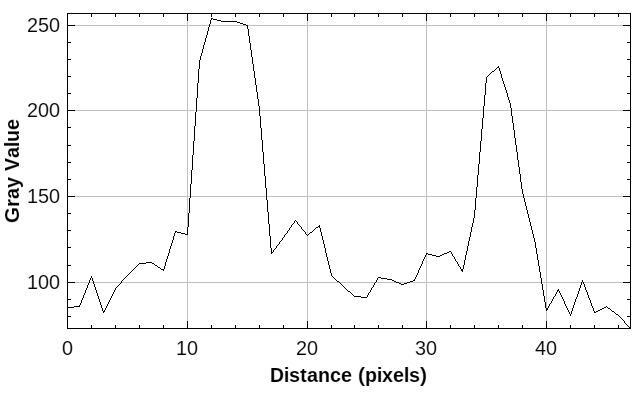}
        \caption{}
        \label{fig:b}
    \end{subfigure}

    \caption{(a) Central spots due to reflection from the display. (b) ImageJ scan along the yellow line.}
    \label{fig:spots}
\end{figure}

When the laser had a slant incidence on the display, we observed another bright spot near the central spot of the diffraction pattern (See  Fig. \ref{fig:spots}). 

\begin{figure}[h!]
    \centering
\input{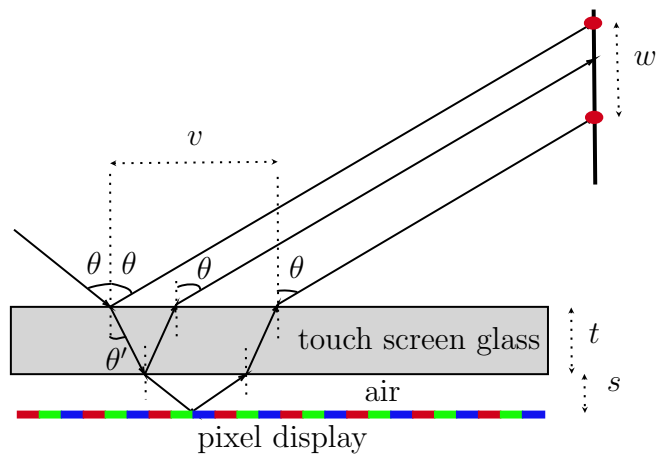}
\captionsetup{justification=raggedright,singlelinecheck=false}
\caption{Schematic ray diagram for the reflection experiment. In the experiment with the Redmi 9 Pro, we observed only two dots on the observation screen. The first dot is due to the reflection from the top of the glass surface. The observation of only two red dots instead of three suggests that there is no air gap between the glass and the pixel display. This absence of an air gap would cause the reflections from the bottom of the glass surface and the pixel display to merge, resulting in a single dot rather than two separate ones.}
    \label{fig:ray_diag}
\end{figure}

Notably, the distance of the additional spot from the central one  varies with the angle of incidence of the laser ray on the display. This is because a part of the light from the laser is also reflected by the top surface of the display's glass, while the rest is reflected from the pixel layer beneath. 
The ray diagram depicting this process is presented in Fig.~\ref{fig:ray_diag}, which illustrates how the two spots form on the observation screen through the reflection from the top surface of the touch glass and the pixel layer beneath it. Note that the pixel display reflects only a specific polarization of light. Ideally, a glass slab with a reflective surface should give multiple bright spots on an observation screen \cite{AhmetTPT,Spot_Arnab}, but in our case, this is limited by the intensity of the laser light and a less reflective surface of the touch glass of the phone's display. 

Figure \ref{fig:ray_diag} shows a ray of light partially reflecting from the top and bottom surfaces of the touch glass, and the pixel display, which is positioned at a distance $s$ away from the bottom of the glass. Let the thickness of the glass be $t$. From Fig.~\ref{fig:ray_diag} we see 
\begin{align}
    y=2(t\tan{\theta'}+s\tan{\theta}),
\end{align}
where $\theta'$ is the angle of refraction going from air to glass.
From Fig.~\ref{fig:ray_diag} we can also see that
\begin{align}
    \cot{\theta}=\frac{w}{{y}}.
\end{align}
Using Snell's law, $\mu_{a}\sin{\theta}=\mu_{g}\sin{\theta'}$, where $\mu_a$, and $\mu_g$ are refractive indices of air and glass respectively.  The above equation then can be written as 
\begin{align}
   \frac{w}{2}=\frac{\mu_a\sqrt{1-\sin^2\theta}}{\mu_g\sqrt{1-\left(\frac{\mu_a}{\mu_g}\right)^{2}\sin^2\theta}}t+s \label{eq:w}.
\end{align}
For the incident angle \(\theta\), the equation governing \(w\) is expressed as:
\begin{align}
    \frac{w}{2}=\eta(\theta) t +s,
\end{align}
where \(\eta(\theta)\) is the coefficient of $t$ of Eq. (\ref{eq:w}).

The angle \(\theta\) can be systematically varied, and the distance between the spots can be measured by capturing and analyzing an image. The plot of the distances between the spots and \(\eta(\theta)\) exhibits linearity. The slope of this line provides the thickness of the glass, while the intercept denotes the distance $s$ between the glass and the pixel layer beneath it.

\begin{figure}[h!]
    \centering
\includegraphics{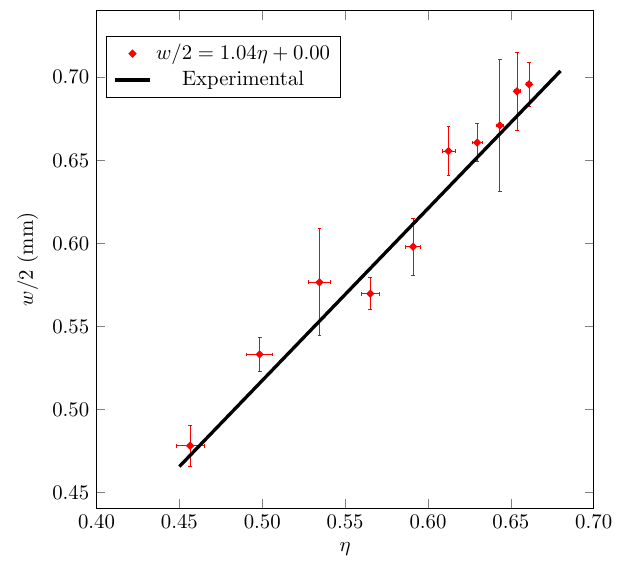}
\caption{Plot of $w/2$ vs $\eta$ in the reflection experiment.}
\label{fig:spot_expt}
\end{figure}

The graph showing the obtained values of \(w\) plotted against \(\eta\) is presented in Fig. \ref{fig:spot_expt}. The best fit line plotted gives the thickness of the touch glass on the utilized phone to be \(1.04\pm0.08\) mm, and the gap between the glass and the pixel display to be \(0.00\pm 0.04\). The uncertainty in this case pertains to the best-fit intercept, and it is one order of magnitude greater than the true value. This suggests that the gap is nearly non-existent, which matches our observation of two spots. These findings will be further corroborated by the upcoming water drop experiment.

\section{Optics of the water-drop-lens} 
\begin{figure}[htbp]
    \centering

    \begin{subfigure}[b]{0.35\textwidth}
        \centering
        \includegraphics[width=\linewidth]{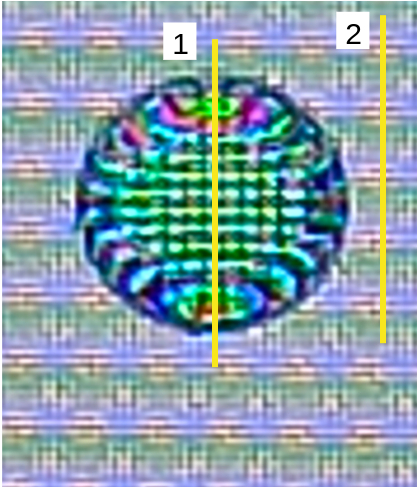}
        \caption{}
        \label{fig:a}
    \end{subfigure}

    \vspace{0.5em} 

    \begin{minipage}{\textwidth}
        \centering
        \begin{subfigure}[b]{0.48\textwidth}
            \centering
            \includegraphics[height=5cm]{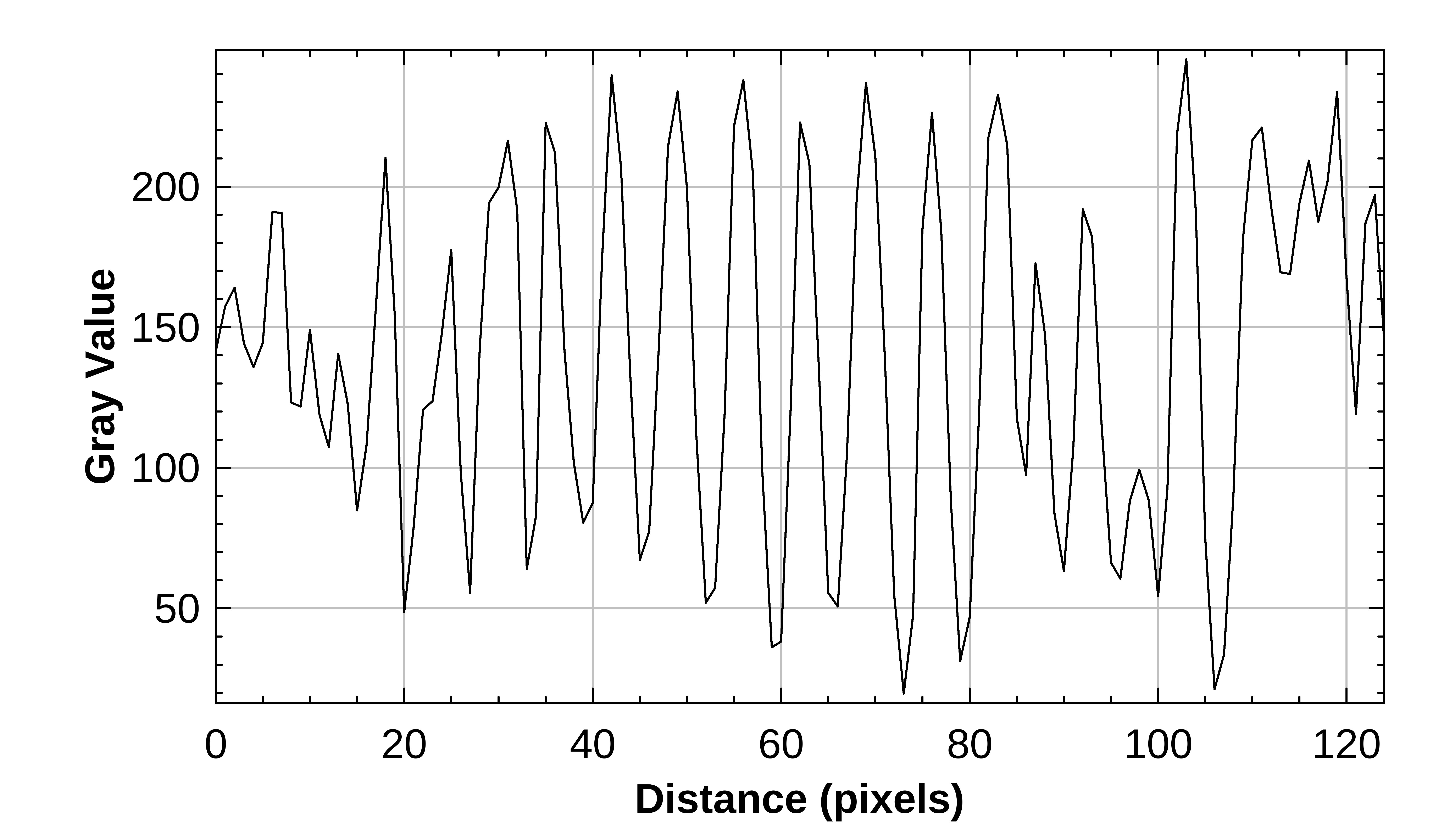}
            \caption{}
            \label{fig:b}
        \end{subfigure}
        \hspace{-0.03\textwidth}
        \begin{subfigure}[b]{0.48\textwidth}
            \centering
            \includegraphics[height=5cm]{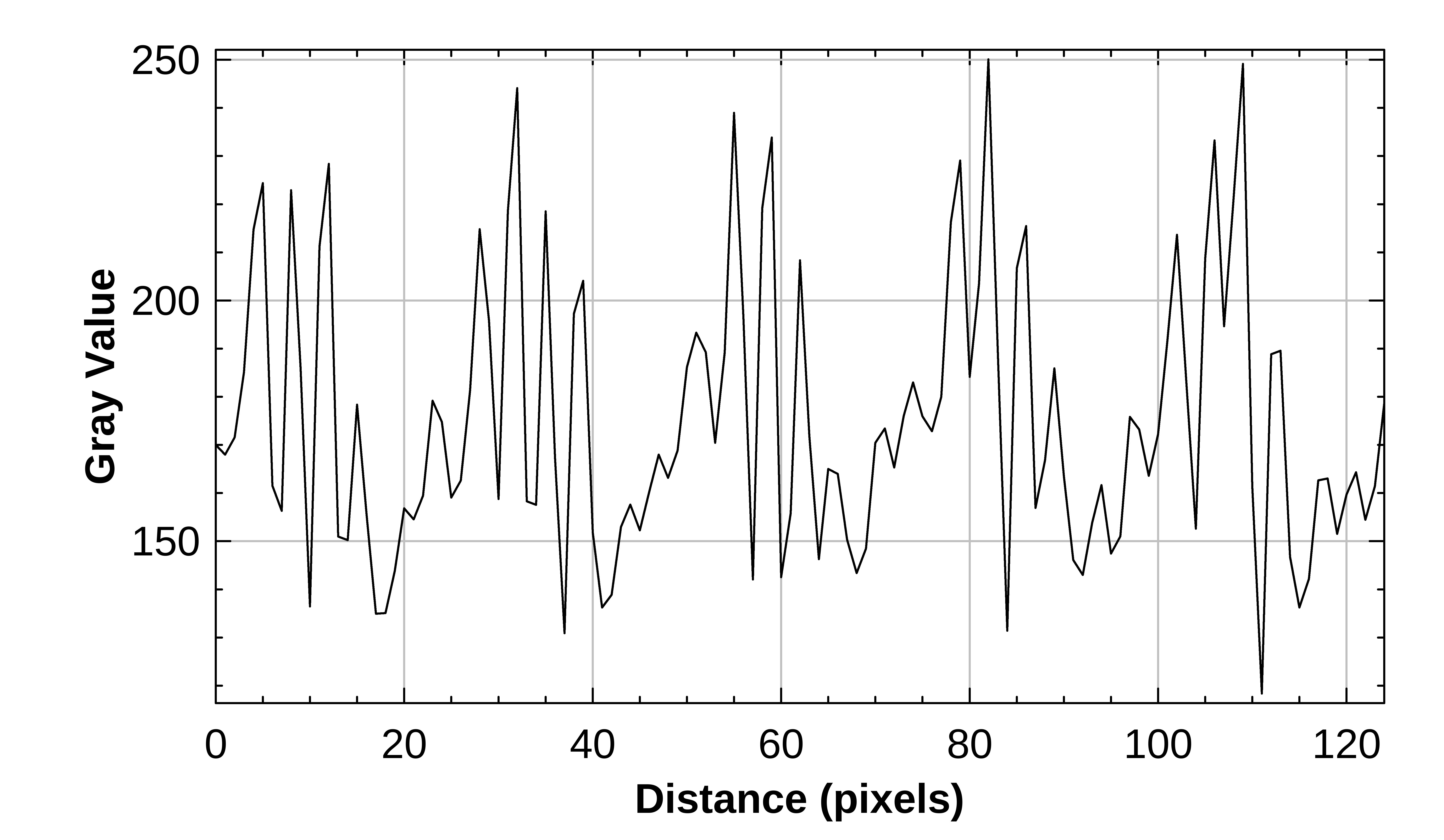}
            \caption{}
            \label{fig:c}
        \end{subfigure}
    \end{minipage}
    \captionsetup{justification=raggedright,singlelinecheck=false}
    \caption{(a) is the drop with magnified pixels. (b) is the ImageJ scan along line 1, which shows magnified pixels. (c) is the ImageJ scan along line 2 on the phone's display to get the PPI of the pixel display.}
    \label{fig:Magnifieddrop}
\end{figure}

When small water drops (~1–4\,mm in diameter) are placed on the touch glass of a bright phone screen, grid-like patterns can be seen through the drops (see Fig. \ref{fig:Magnifieddrop}(a)). These patterns are magnified images of the underlying pixels.\cite{Freeland}

\begin{figure}[h!]
    \centering
    \includegraphics[width=0.5\linewidth]{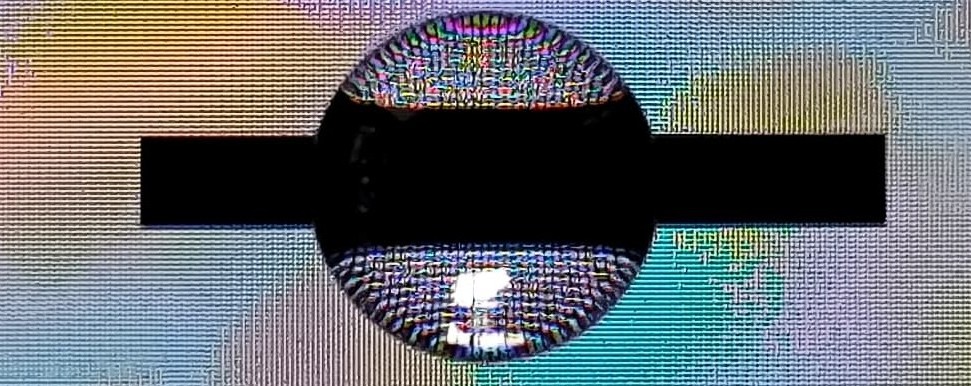}
    \caption{Another method to find magnification for large drops}
    \label{fig:second_method}
\end{figure}

For larger drops, the magnification is very small, making the individual pixels difficult to resolve. However, the magnification is observable
for a water droplet above a black line (such as a letter ``I" or ``L") on a white background. Figure~\ref{fig:second_method} shows an example of this magnification.

\begin{figure}[h!]
\includegraphics[scale=0.3]{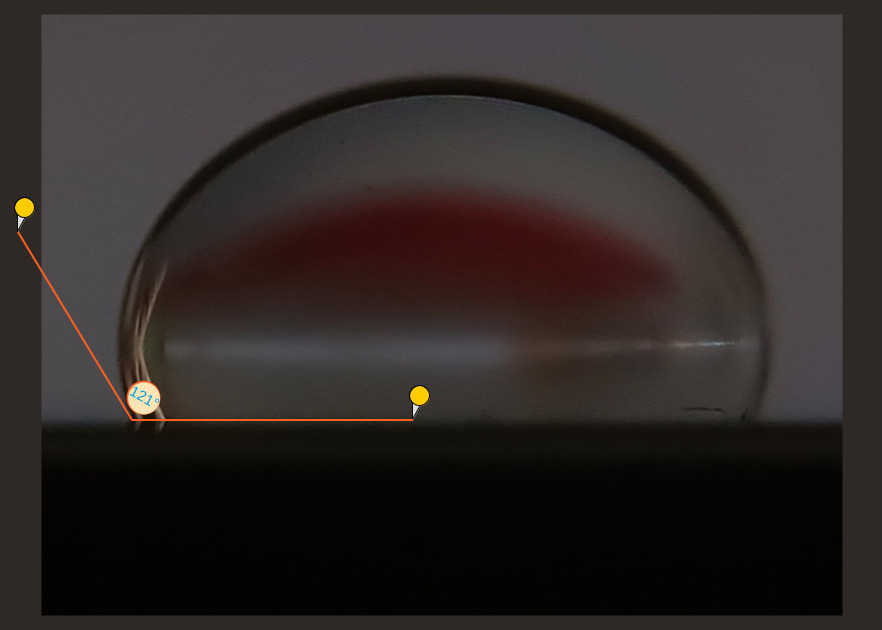}
\captionsetup{justification=raggedright,singlelinecheck=false}
\caption{A waterdrop resting on the phone's display. An online angle measurement tool is used to measure the contact angle greater than $90^\circ$. The online tool measures the angle between the yellow pointers and the centre.}
\label{fig:sideview}
\end{figure}

Freeland et al. \cite{Freeland} modelled the drop as hemispherical and used the thin lens approximation to justify the magnification formula. In the reflection experiment, we determined that the distance of these pixels from the water drop lens ($\approx$1\,mm) is comparable to the radius of the water drop lens. Thus, a more accurate approach would be to refrain from applying the thin lens approximation in this case \cite{Ghatak_thinlens}. Also, Fig.\,\ref{fig:sideview} shows an image of the water drop lens placed on a phone's touch glass, indicating that the drop is not hemispherical. Generally, for small enough drops, surface tension pulls the drop to a spherical sector-like shape that minimizes surface area. As drop size increases, gravitational deformation becomes significant. In general, the gravitational potential energy is of the order of \(\rho g r^4\), while the surface energy is of the order of \(\sigma r^2\), where \(\sigma\) is the surface tension of water. Thus, it is expected, since the typical water droplets used have radii of the order of or greater than \(\sqrt{\frac{\sigma}{\rho g}}\approx 2.7 \text{ mm}\), they will experience significant deformation due to gravity. 

\begin{figure}[h!]
\input{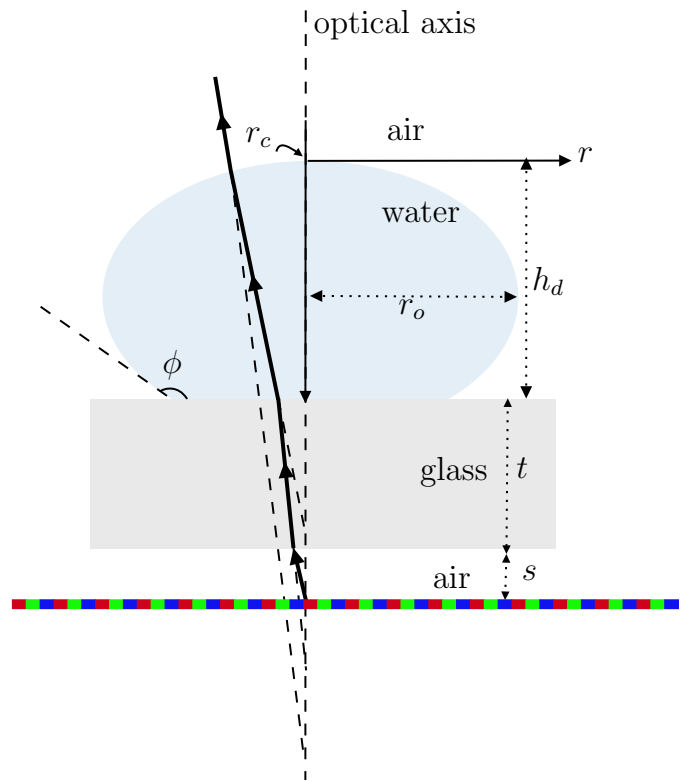}
\captionsetup{justification=raggedright,singlelinecheck=false}
\caption{Schematic diagram illustrating a water drop lens resting on the display, depicting the length parameters employed in the theoretical calculations. Here $r_c$ is the radius of the curvature at the top of the drop, $r_o$ is the maximum radius of the drop, and $h_d$ is the height of the drop.}
\label{fig:drop_parameters}
\end{figure}

Keeping these aspects in mind, we solve the problem by explicitly invoking Snell's law at all the interfaces and considering a general shape of the drop. Figure \ref{fig:drop_parameters} depicts the schematic diagram of the setup.

\begin{figure}[h!]
    \centering
\input{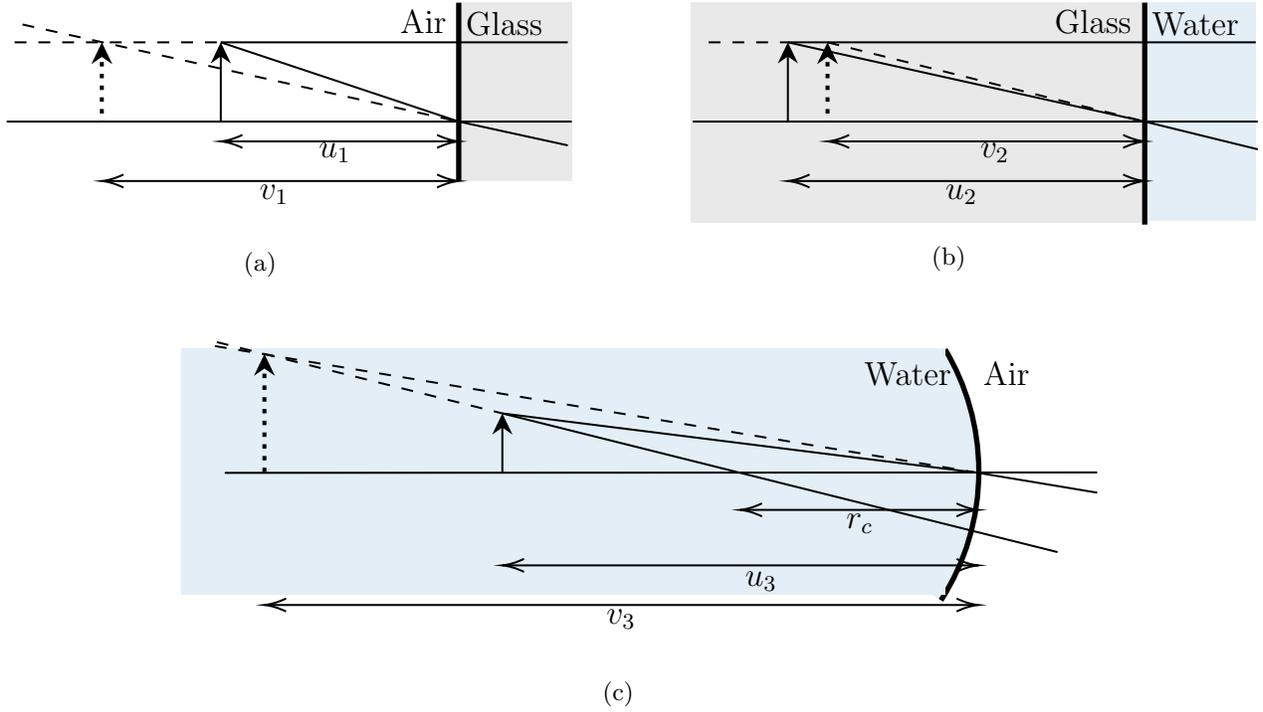}
\captionsetup{justification=raggedright,singlelinecheck=false}
    \caption{Ray diagrams for the refractions at three interfaces: (a) air-glass, (b) glass-water, (c) water-air. Solid upright arrows represent the object, while the solid lines represent the rays. The rays are extended backwards, shown as dashed lines, to obtain the virtual images, depicted by dotted arrows.}
    \label{fig:enter-label}
\end{figure}

Starting from the arrays of pixels, there are three refractions involved: (i) air-glass interface, (ii) glass-water interface, and (iii) water-air interface. We will perform explicit calculations for each of the interfaces. The refractive indices for air, water, and glass are $\mu_a = 1.00$, $\mu_w = 1.33$, and  $\mu_g = 1.50$, respectively.\cite{corningGG5} 

\textbf{Image formation through air-glass interface:}

Figure~\ref{fig:enter-label}(a) shows the object (pixels in the setup) in air. The image after the first refraction can be constructed as shown. The ray passing normal to the surface is undeflected. Since the surface is flat, the normal is always parallel to the optic axis. This ray in the ray diagram shows that the magnification of the image is 
\begin{equation}
    \label{M1}
    M_1 = 1\,.
\end{equation}
The ray passing through the pole is refracted, with the optic axis (see Fig. \ref{fig:drop_parameters}) serving as the normal. Applying Snell’s law with the small-angle approximation now yields
\begin{equation}
    \frac{\mu_a}{u_1} = \frac{\mu_g M_1}{v_1},
\end{equation}
where \(u_1\) and \(v_1\) are the object and image distances for this refraction.
Thus, we get
\begin{equation}
    \label{img1}
    v_1 = \frac{\mu_g}{\mu_a} u_1 \,.
\end{equation}

\textbf{Image formation through glass-water interface:}

The image formed from the first refraction is a secondary object for the glass-water interface. Figure ~\ref{fig:enter-label}(b) presents a schematic diagram illustrating the image construction through this interface. Once more, the ray passing normal to the surface remains undeflected, yielding the magnification of the second image as
\begin{equation}
    \label{M2}
    M_2 = 1 \,.
\end{equation}
Similar to the first interface, the ray passing through the pole undergoes refraction, with the optic axis serving as the normal. Applying Snell’s law with the small-angle approximation, we now obtain
\begin{equation}
    \frac{\mu_g}{u_2} = \frac{\mu_w M_2}{v_2},
\end{equation}
where \(u_2\) and \(v_2\) are the object and image distances for this refraction.
Thus, we get
\begin{equation}
    \label{img2}
    v_2 = \frac{\mu_w u_2}{\mu_g}\,.
\end{equation}

\textbf{Image formation through the water-air interface:}

The third refraction occurs at the curved surface between water and air. Image construction for this interface is illustrated in (Fig.~\ref{fig:enter-label}(c)). The image formed by the previous interface serves as an object for this refraction. Here, the ray passing through the centre of curvature at the top of the drop travels normal to the refracting surface and remains undeflected. This leads to the equation

\begin{equation}
\frac{1}{u_3 - r_c} = \frac{M_3}{v_3 - r_c},
\end{equation}
where \(u_3\) and \(v_3\) are the object and image distances for this refraction.

Subsequently, the ray passing through the pole undergoes refraction, with the optic axis serving as the normal. Utilizing the small-angle approximation, we obtain
\begin{equation}
    \frac{\mu_w}{u_3} = \frac{\mu_a M_3}{v_3}\,.
\end{equation}
From these equations, we get
\begin{equation}
    \label{M3}
    M_3 = \frac{\mu_w r_c}{\mu_a u_3 - \mu_w u_3 + \mu_w r_c}\,,
\end{equation}
\begin{equation}
    \label{img3}
    v_3 = \frac{\mu_a u_3 r_c}{\mu_a u_3 - \mu_w u_3 + \mu_w r_c}\,.
\end{equation}

Examining the successive refractions in our setup, depicted in Fig.~\ref{fig:drop_parameters}, we observe that
\begin{align}
    \label{r1} u_1 &= s,\\
    \label{r2} u_2 &= t+v_1, \\
    \label{r3} u_3 &= h_d+v_2,
\end{align}
where \(s\) is the size of the airgap, \(t\) is the thickness of glass and \(h_d\) is the height of the drop. The total magnification is $M =M_1M_2M_3$. Thus, using Eqs. (\ref{M1}), (\ref{img1}), (\ref{M2}), (\ref{img2}), (\ref{M3}), (\ref{r1})-(\ref{r3}), we obtain \(M\) in terms of \(s\), \(t\), \(h_d\) and \(r_c\) as
\begin{equation}
    M = \frac{\mu_w r_c}{\mu_w r_c - (\mu_w - \mu_a)\left(h_d + \frac{\mu_w}{\mu_g}t + \mu_w s\right)}\,.
\end{equation}

A neater way to write this expression is
\begin{equation}
    \frac{1}{M} = 1 - \left(1-\frac{\mu_a}{\mu_w}\right)\frac{1}{r_c}\left(h_d + \frac{\mu_w}{\mu_g}t + \mu_w s\right)\,.
\end{equation}
Using the numerical values of the refractive indices, we have
\begin{equation}
    \label{eq:mag_Rc}
    \frac{1}{M} = 1 - \frac{1}{3r_c}\left(s + \frac{2}{3}t + \frac{3}{4}h_d\right)\,.
\end{equation}

Using the numerical analysis of the Appendix, one can write $r_c, h_d$ in terms of the contact angle $\phi$ and the maximum radius of the drop $r_o$ (see Fig. \ref{fig:drop_parameters}). Since the contact angle is a constant, Eq. (\ref{eq:mag_Rc}) is an implicit function of \(r_o\), which is the measurable parameter in this experiment. Thus, this function describes how the magnification measured \(M\) should vary with the varying size of the drops \(r_o\).

If the drops are small, with \(r_o \ll \sqrt{\sigma/\rho g}\), gravity may be neglected against surface tension and one may consider the drop-air interface to be spherical, in which case \(r_c \rightarrow r_o\) and \(h_d \rightarrow r_o(1-\cos{\phi})\). Then, Eq. (\ref{eq:mag_Rc}) reduces to
\begin{align}
    \frac{1}{M}=\frac{3+\cos{\phi}}{4}-\frac{1}{3r_o}\bigg(s+\frac{2}{3}t\bigg)\,.\label{eq:Spherical_approximation}
\end{align}
For $\phi=90^\circ$, the above equation resembles Eq.~(4) of Ref. \cite{Freeland} but with a different intercept value. This difference can be attributed to our adoption of a more comprehensive approach instead of the thin lens approximation utilized in Ref.\cite{Freeland}.

\begin{figure}[h!]
    \centering
\includegraphics{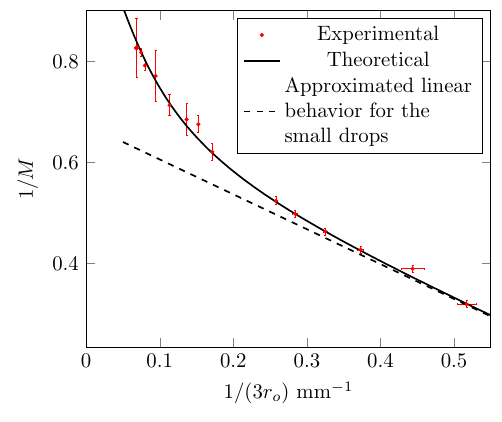}
\caption{plot of $1/M$ vs $1/(3r_o)$ for the same phone. The $\phi$ is 115$^\circ$ and $t$ is 0.98\,mm.}
\label{fig:drop_expt}
\end{figure}

The experimental procedure is as follows: a picture of the magnified pixel through a water-drop is taken along with a scale for calibration. The image is then processed in ImageJ to measure the magnified pixels. The actual pixel values are obtained from the technical specs of Redmi Note 9 Pro (391 PPI \cite{note9pro}), which is cross-verified by scanning the area outside the drop (see Fig. \ref{fig:Magnifieddrop}(c)). The magnification is measured with these values, and a graph of the inverse of magnification vs \(1/(3r_o)\) is plotted. Equation (\ref{eq:Spherical_approximation}) predicts an approximately linear curve for small drops. As seen in Fig. \ref{fig:drop_expt}, we were able to measure a departure from linear behavior for sufficiently large water drop radii.

In Appendix \ref{sec:numerical}, we obtain the differential equation for the curve \(r(h)\) representing the shape of the drop by balancing the force due to gravity, the surface tension of the water drop, and hydrostatic pressure. The equation can be solved numerically to get a set of values \{\(r_c,h_d,r_o,\phi\)\}. We use the experimentally obtained value of \(r_o\) to determine the values of \(r_c\) and \(h_d\) for generating a theoretical curve between \(1/M\) and \(1/3r_o\) as per Eq. (\ref{eq:mag_Rc}).
 In principle, we could calculate different curves for different contact angle $\phi$ values and \(s+(2/3)t\) values. For our calculations, we used the glass thickness  \(t\) and air gap \(s\) values observed in the reflection experiment. We experimentally obtained the contact angle by taking a side-view photograph of the drop. This image serves two purposes. First, it confirms that the drop is non-hemispherical, justifying the need for a general calculation. Second, it allows us to get the contact angle using an online protractor \cite{Protractor}. Figure \ref{fig:sideview} displays the side view of one of the water droplets with a measured angle value of 121$^\circ$. We analyzed a total of 10 drops, where the angle $\phi$ was found to be 117$\pm 4^\circ$.

Figure \ref{fig:drop_expt} displays the experimental data along with the theoretical curve. The theoretical curve is plotted for a contact angle of $\phi=120^\circ$ and $s+(2/3)t=0.67$. The value of $\phi$ was chosen first using the linear trend noted for small drops, and then varied by hand until the curve fit looked good enough. This fit by visual inspection was seen to fall within the experimental range.

\section{Conclusion}
Our exploration of smartphone display optics unveils intriguing connections between diffraction patterns, reflections, and lens behavior. The diffraction experiment enables teachers and students to measure the PPI of their phone and demonstrate the smartphone display's potential as an optical grating. In some modern phones, the angled arrangement of pixels creates an X-shaped diffraction pattern, offering a unique challenge for students to explore. Reflection studies not only revealed the thickness of the touch glass and potential air gap but also highlighted the pixel display's properties. Considering the curved shape and Snell's law, the water drop experiment yielded experimental data consistent with our theoretical predictions. Auxiliary measurements of contact angles and characteristic parameters underscored the reliability of our methodologies. In principle, all three experiments can be performed on any other smartphone.  Overall, these experiments provide accessible and
 educational insights into the hidden optical features of smartphone displays. These insights are valuable for both scientific exploration and educational demonstrations.

\section*{Acknowledgement}
We acknowledge the support of the Olympiad program of the Government of India, Department of Atomic Energy, under project identification number RTI4001. We thank Charudatt Kadolkar
 for the helpful discussions.

\newpage 

\appendix

\section{The shape of a drop}\label{sec:numerical}

\begin{figure}[h!]
    \centering
    \includegraphics[scale=0.2]{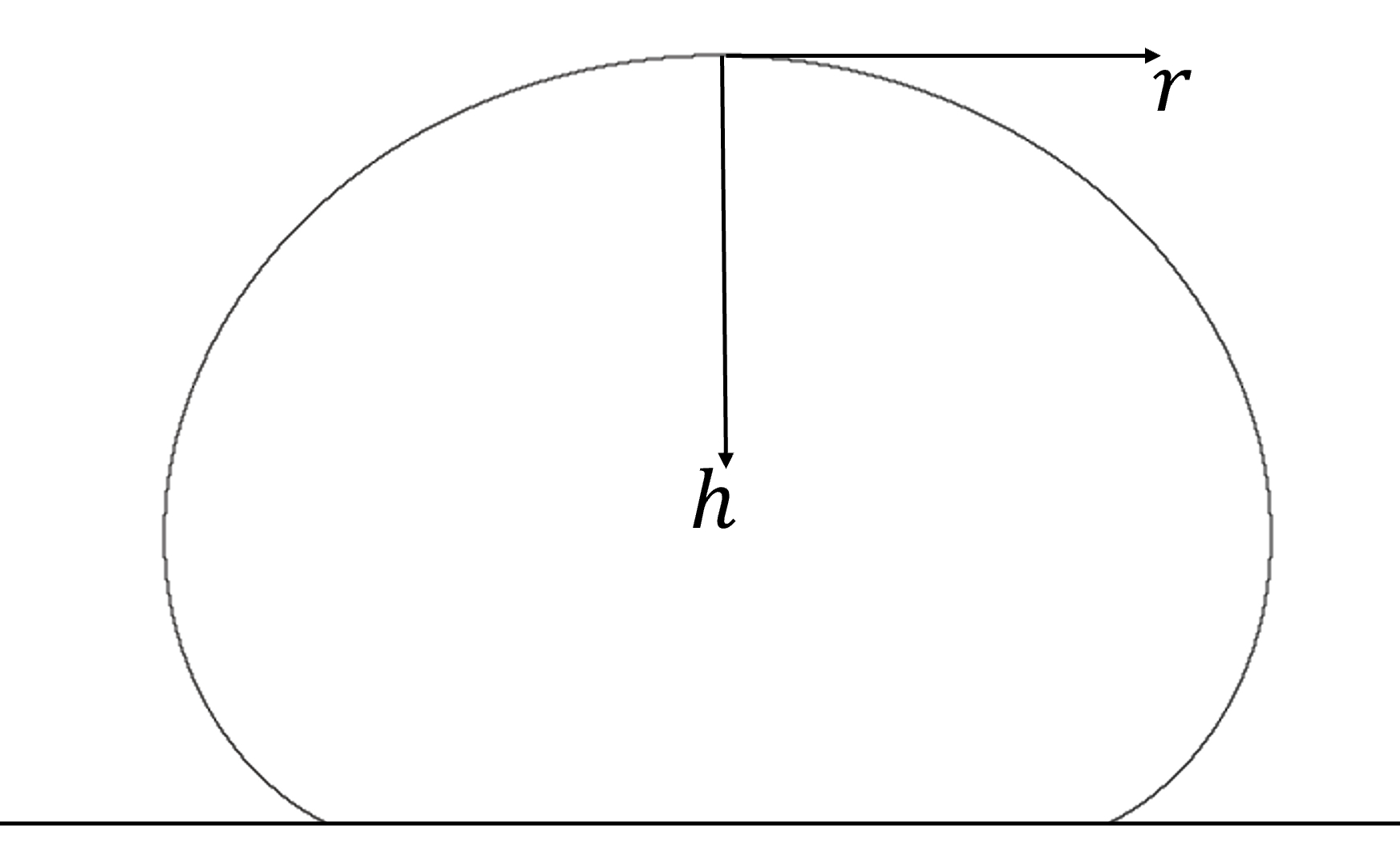}
    \caption{Drop figure produced by computation}
    \label{fig:Typical drop}
\end{figure}

Consider a drop resting on a surface. The surface should be non-wetting, i.e. such that the contact angle of the liquid with it is greater than \(90^\circ\). In the case of water, such surfaces are called hydrophobic surfaces. The key forces at play are the surface tension of water  holding the drop-together \cite{berry1971} and gravity trying to pull the water downward. In equilibrium, we wind up with an azimuthally symmetric shape, as shown in Fig.\,\ref{fig:drop_parameters}. We construct axes ($h,r$) as shown in Fig.~\ref{fig:Typical drop}, with the origin at the topmost point of the drop. Thus, given the azimuthal symmetry, finding the shape of the drop is equivalent to finding the correct function \(r(h)\). 

First, consider the balance of forces inside the drop. The only agents at play are pressure and gravity. Let \(\rho\) be the density of liquid and \(g\) be the acceleration due to gravity. Equating the net force per unit volume inside the fluid to zero, we see that the pressure inside the fluid must obey
\begin{equation}
    -\vec{\nabla} P + \rho \vec{g} = 0,
\end{equation}
which gives the pressure as
\begin{equation}
    P=P_c+\rho g h \label{Ph},
\end{equation}
where \(P_c\) is the pressure just inside the drop at $r=0$.

Looking at the surface element at the origin, equilibrium under surface tension means the drop is flat at the top. Hence, for a given radius of curvature \(r_c\) of the surface element at the origin, equating the force per unit area on it to zero gives
\begin{equation}
\label{Pc}
    P_c=P_0+\frac{2\sigma}{r_c},
\end{equation}
where \(P_0\) is the external pressure and \(\sigma\) is the surface tension.

\begin{figure}[h]
    \centering
    \includegraphics[scale=0.2]{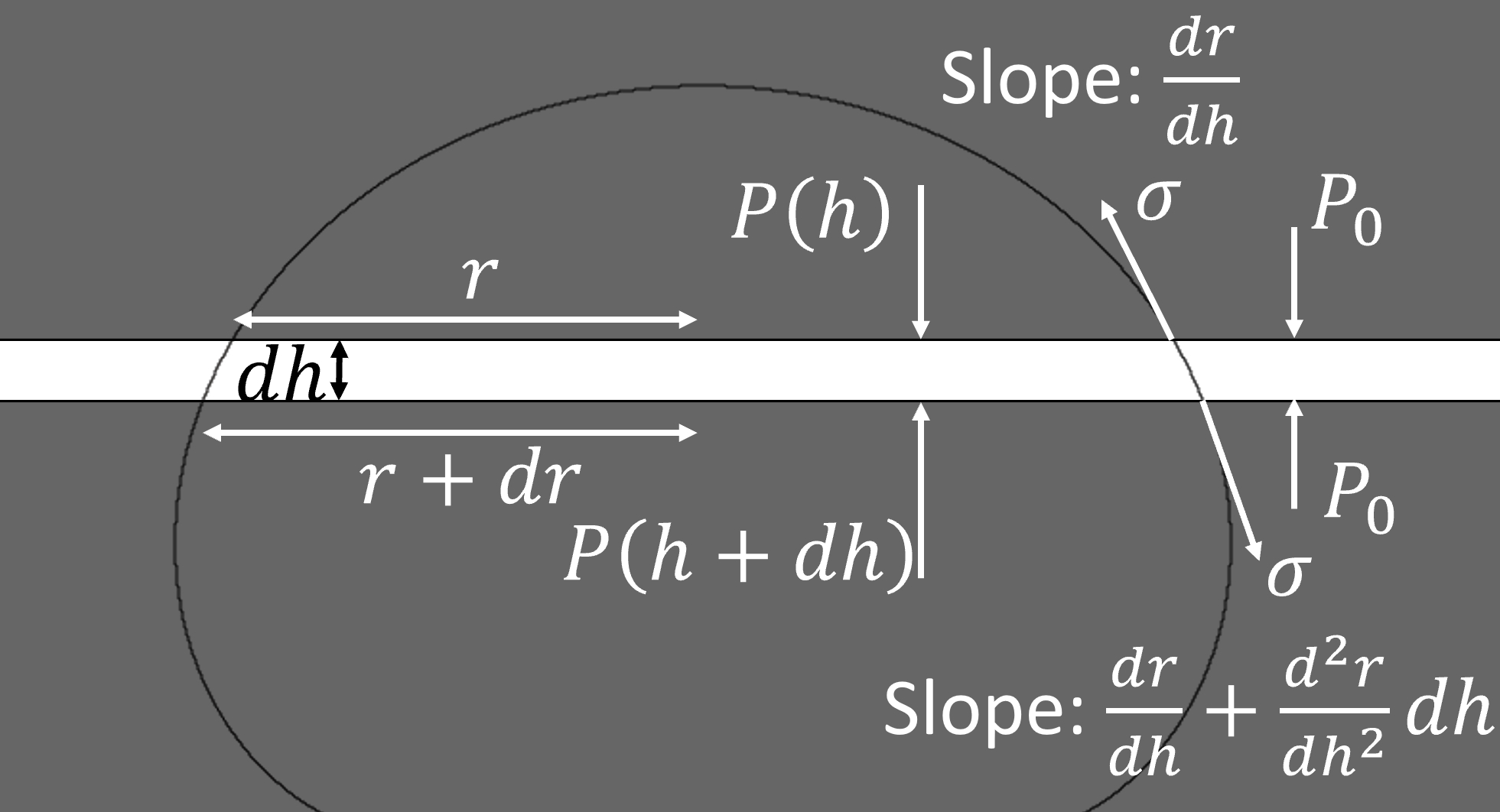}
    \caption{Surface tension and pressure acting on a horizontal element.}
    \label{fig:horizontalelement}
\end{figure}

We now balance the forces acting on a horizontal slice of the drop of thickness \(dh\). The forces due to pressure and surface tension (see Fig. \ref{fig:horizontalelement}) must balance the gravitational pull on the element.

The upward force due to pressure is
\begin{align}
    F_P & =(P(h+dh)-P_0)\times \pi (r+dr)^2-(P(h)-P_0)\times \pi r^2\\
    \Rightarrow F_P & =\bigg(\frac{dP}{dh}\times \pi r^2 +(P-P_0)\times 2\pi r \frac{dr}{dh}\bigg)dh .
\end{align}
The upward force due to surface tension is
\begin{align}
    F_\sigma & = -d\bigg(2\pi r \sigma \frac{1}{\sqrt{1+\big(\frac{dr}{dh}\big)^2}}\bigg)\\
    \Rightarrow F_\sigma & = -2\pi\sigma\Bigg(\frac{\frac{dr}{dh}}{\sqrt{1+\big(\frac{dr}{dh}\big)^2}}-\frac{r\frac{dr}{dh}\frac{d^2r}{dh^2}}{\Big(1+\big(\frac{dr}{dh}\big)^2\Big)^{3/2}}\Bigg)dh.
\end{align}

These forces balance the downward force of gravity, equal to \(\pi r^2 \rho g\,dh\). Thus, we have, by substituting for pressure from Eq. (\ref{Ph}) and Eq. (\ref{Pc}),
\begin{equation}
    \bigg(\rho g\times \pi r^2 +\bigg(\frac{2\sigma}{r_c}+\rho gh\bigg)\times 2\pi r \frac{dr}{dh}\bigg)dh-2\pi\sigma\frac{\frac{dr}{dh}}{\sqrt{1+\big(\frac{dr}{dh}\big)^2}}\Bigg(1-\frac{r\frac{d^2r}{dh^2}}{1+\big(\frac{dr}{dh}\big)^2}\Bigg)dh=\pi r^2 \rho g dh.
\end{equation}
This simplifies to
\begin{equation}
    \label{Drop}
    \frac{2r}{r_c}+\frac{\rho g}{\sigma}hr=\frac{\frac{dr}{dh}}{\sqrt{1+\big(\frac{dr}{dh}\big)^2}}\Bigg(1-\frac{r\frac{d^2r}{dh^2}}{1+\big(\frac{dr}{dh}\big)^2}\Bigg).
\end{equation}
This is the equation determining the shape of the drop and has a dependency on the forces due to the surface tension and gravity.

\begin{figure}[h]
    \centering
    \includegraphics[scale=0.75]{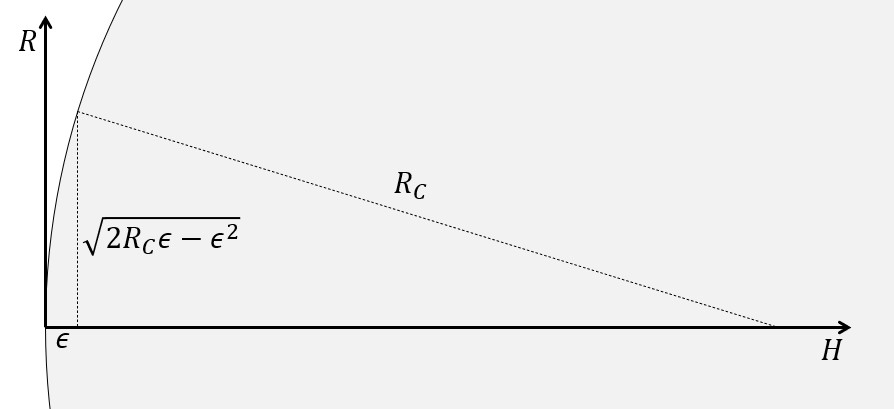}
    \caption{Starting conditions for computation}
    \label{fig:sc}
\end{figure}

The resulting differential equation is solved numerically for the shape of the drop. To facilitate numerical computations, we introduce dimensionless variables and transform the equation. Making substitutions   \(r=\frac{R}{\sqrt{\rho g/\sigma}}\)and \(h=\frac{H}{\sqrt{\rho g/\sigma}}\) gives us,
\begin{equation}
    \frac{2R}{R_C}+HR=\frac{\frac{dR}{dH}}{\sqrt{1+\big(\frac{dR}{dH}\big)^2}}\Bigg(1-\frac{R\frac{d^2R}{dH^2}}{1+\big(\frac{dR}{dH}\big)^2}\Bigg),
\end{equation}
with \(r_c=R_C/\sqrt{\rho g/\sigma}\). Starting from a point close to the origin where \(R \rightarrow 0\) and \(\frac{dR}{dH} \rightarrow \infty\), we incrementally compute the shape of the drop. Since we know the dimensionless radius of curvature to be \(R_C\), the starting point \(H=\epsilon\) will have \(R=\sqrt{2R_C\epsilon-\epsilon^2}\) and \(dR/dH=(R_C-\epsilon)/\sqrt{2R_C\epsilon-\epsilon^2}\) (See Fig.\,\ref{fig:sc}). The shape of the drop is obtained by computing further with small increments on \(H\) up to the point where \(dR/dH\) drops below \(\cot{\theta}\), where \(\theta\) is the contact angle. That is the point at which the drop will rest on the surface. Thus, we can obtain R(H), for different values of \(R_C\) as seen in simple C code. Figure \ref{fig:Typical drop} has been produced by such a computation.

The code yields the shape of the drop, \(R(H)\), however, this is not important for further analysis. We only need the numerical relationship between three quantities: $R_C$, $H_D$,  and \(R_O\). Here, \(R_o\) is the dimensionless measures of \(r_o\), the maximum value of \(r\), which is the observed radius of the drop when seen from the top; and \(H_D\) is the value of \(H\) up to where the computation is done, which is a dimensionless measure of the height of the drop.

\end{document}